\documentclass[conference]{IEEEtran}

\usepackage{balance}  %
\usepackage{graphicx,epsfig}
\usepackage{mathrsfs}
\usepackage[font=footnotesize,caption=false]{subfig}
\usepackage[dvipsnames]{xcolor}
\usepackage{lipsum}
\usepackage{amsmath,amssymb,xspace,amsthm}
\usepackage{url}
\usepackage{multirow}
\usepackage[]{cite}
\usepackage{booktabs}
\graphicspath{{fig/}}

\newtheoremstyle{hypstyle}{\topsep}{\topsep}{\itshape}{}{\bfseries}{:}{.5em}{}
\theoremstyle{hypstyle}

\usepackage[ruled,vlined]{algorithm2e}
\SetArgSty{textrm}
\SetFuncSty{textsc}
\SetKwFunction{algo}{HogePiyo}
\SetKwFunction{proc}{proc}
\SetKwProg{Procedure}{Procedure}{}{}
\SetKwProg{Function}{Function}{}{}

\definecolor{mycommcolor}{rgb}{0.1,0.1,0.45} %

\SetCommentSty{mycommfont}

\makeatletter
\newcommand{\algrule}[1][.2pt]{\par\vskip.5\baselineskip\hrule height #1\par\vskip.5\baselineskip}
\makeatother

\newcommand{\myfnsty}[1]{\textit{#1}}

\newcommand{\myparagraphF}[1]{\noindent {\bf #1.}}
\newcommand{\myparagraph}[1]{\vspace{1em} \noindent {\bf #1.}}
\newcommand{\etal}[1]{{\it et al.}}
\definecolor{lipsumcolor}{rgb}{0.7,0.7,0.7}

\newcommand{\abs}[1]{\left| #1 \right|}
\newcommand{\set}[1]{\left\{ #1 \right\}}
\newcommand{\ignore}[1]{}

\newcommand{\secit}{}

\let\oldenumerate\enumerate
\renewcommand{\enumerate}{
   \oldenumerate
   \setlength{\itemsep}{4pt}
   \setlength{\parskip}{0pt}
   \setlength{\parsep}{0pt}
}
\let\olditemize\itemize
\renewcommand{\itemize}{
   \olditemize
   \setlength{\itemsep}{4pt}
   \setlength{\parskip}{0pt}
   \setlength{\parsep}{0pt}
}

\begin{document}

\IEEEoverridecommandlockouts

\title{Cut Tree Construction from Massive Graphs$^\ast$
\thanks{$^\ast~$This work is done while all authors were at National Institute of Informatics.
A shorter version of this paper appeared in the proceedings of
ICDM 2016~\cite{short}.}}

\author{\IEEEauthorblockN{
Takuya Akiba$^{\dagger}$ \quad
Yoichi Iwata$^{\ddag}$ \quad
Yosuke Sameshima$^{\#}$ \quad
Naoto Mizuno$^{\S}$ \quad
Yosuke Yano$^{\#}$}

\fontsize{10}{10}\selectfont\itshape
$^{\dagger}$\,Preferred Networks, Inc. \;
$^{\ddag}$\,National Institute of Informatics \;
$^{\#}$\,Recruit Holdings Co., Ltd. \;
$^{\S}$\,The University of Tokyo \; \\

\fontsize{9}{9}\selectfont\ttfamily\upshape
akiba@preferred.jp,
yiwata@nii.ac.jp, \\
yosukes@indeed.com,
mizuno@eps.s.u-tokyo.ac.jp,
yosukey@indeed.com
}

\maketitle

\begin{abstract}
  The construction of \emph{cut trees} (also known as \emph{Gomory-Hu trees})
  for a given graph enables the minimum-cut size of the original graph to be obtained for any pair of vertices.
  Cut trees are a powerful back-end for graph management and mining,
  as they support various procedures related to the minimum cut, maximum flow, and connectivity.
  However, the crucial drawback with cut trees is the computational cost of their construction.
  In theory, a cut tree is built
  by applying a maximum flow algorithm for $n$ times,
  where $n$ is the number of vertices.
  Therefore, naive implementations of this approach result in
  cubic time complexity, which is obviously too slow for
  today's large-scale graphs.
  To address this issue, in the present study,
  we propose a new cut-tree construction algorithm tailored to real-world networks.
  Using a series of experiments, we demonstrate that
  the proposed algorithm is several orders of magnitude faster than previous algorithms
  and it can construct cut trees for billion-scale graphs.
\end{abstract}

\section{Introduction}
\label{sec:introduction}

The \emph{minimum cut} (\emph{min-cut}), \emph{maximum flow} (\emph{max-flow}), and \emph{connectivity} are fundamental concepts in graph theory.
For a pair of vertices $s$ and $t$,
the $s$-$t$ min-cut is the minimum set of edges
such that the removal of any one edge makes $s$ and $t$ disconnected.
The $s$-$t$ max-flow is the
flow from $s$ to $t$ with the maximum amount
(see Section~\ref{sec:preliminaries} for a formal definition).
The beautiful mathematical duality
of the \emph{min-cut max-flow theorem}~\cite{elias1956note,ford1956maximal}
states that the values of the $s$-$t$ min-cut and the $s$-$t$ max-flow are equal.
This value is also called the \emph{connectivity} between $s$ and $t$.
As graph-theoretic building blocks,
the min-cut, max-flow, and connectivity are used in
a wide range of areas, including graph analysis and mining~\cite{mkecs/edbt12,mkecs/sigmod13,mkecs/cikm13,goldberg1984finding,Qin2015,Tsourakakis2015,asano2006mining}.

Because of their importance and rich mathematical properties,
a myriad of algorithms for computing the max-flow and min-cut
have been proposed~\cite{Orlin2013,1970:din,DBLP:journals/jacm/GoldbergT88,DBLP:conf/esa/GoldbergHKTW11}.
However, they each have at least quadratic time complexity in theory~\cite{Orlin2013},
making it time consuming to compute
the max-flow between a pair of vertices.
Moreover, practical applications often require the
repeated computation of max-flows for different vertex pairs.
Therefore, the scalability of
connectivity-based network-analysis methods
is severely limited.

However, a graph has \emph{cut trees}~\cite{cut-tree/gomory-hu} (also known as \emph{Gomory--Hu trees}),
which are a succinct encoding scheme of all the min-cuts of the original graph.
In other words, the min-cut of the original graph
can be quickly obtained from the cut tree for any pair of vertices.
Moreover, cut trees are compact,
having a space complexity that is linear with respect to the number of vertices
(see Section~\ref{sec:preliminaries} for a formal definition).

Thus, it appears that cut trees could play a key role
as a powerful back-end for various network-analysis methods.
However, the crucial drawback is the huge computational cost of constructing cut trees.
In general, a cut tree is built by running a max-flow algorithm $n$ times,
where $n$ is the number of vertices~\cite{cut-tree/gomory-hu,cut-tree/gusfield}.
Therefore, naive implementations of this approach have
at least cubic time complexity, which is obviously too slow for
today's large-scale graphs.

\myparagraph{Contributions}
To address the abovementioned issue,
we propose a new cut-tree construction algorithm tailored to real-world networks of interest,
i.e., large-scale social and web graphs.
The proposed algorithm combines a number of new techniques within three main components.
\begin{itemize}
\item First, we aggressively reduce the given graph into smaller graphs using a series of rules,
  allowing the total cut tree for the original graph to be easily obtained from the cut trees for these smaller graphs (Section~\ref{sec:reduce}).
\item Second, to reduce the number of executions of the max-flow algorithm,
  we propose two efficient heuristics to find ``easy'' min-cuts a priori
  (Sections~\ref{sec:packing} and \ref{sec:goal}).
\item Third, to further reduce the time consumption of the max-flow algorithms,
  we discuss the most suitable techniques for these real networks,
  and propose a practical improvement using bidirectional searches (Section~\ref{sec:flow}).
  We also discuss \emph{source--sink pair ordering strategies} (Section \ref{sec:select}).
\end{itemize}

Experimental results using real large-scale networks confirm that the combination of these new techniques yields a highly scalable cut-tree construction algorithm.
Specifically, whereas previous sophisticated implementations could not construct cut trees for graphs with over one million edges in less than ten hours,
the proposed algorithm successfully constructs cut trees
for very large social and web graphs with more than one billion edges within eight hours.
We also confirm that the data size of the cut trees is sufficiently smaller than the original graph itself,
and find that the average query time for the min-cut size is several microseconds.
Overall, our experimental results
verify that the proposed algorithm makes cut trees
a practical back-end for large-scale graph management and mining.

\myparagraph{Applications}
Let us consider some applications of cut trees
that will be enabled by the proposed algorithm.

\begin{itemize}
\item \textit{Application 1}:
  For any two vertices, we can consider their connectivity as an indicator of the strength of the relationship.
  Thus, it is natural to use the connectivity as a feature of prediction tasks related to vertex pairs
  (e.g., the link prediction problem~\cite{liben2007link}).
  Cut trees enable the connectivity to be used for such tasks,
  as the connectivity will be computed for many vertex pairs during the training and evaluation stages.

\item \textit{Application 2}:
  As mentioned above, the min-cut, max-flow, and connectivity are used
  as graph-theoretic building blocks in various graph analysis and mining techniques.
  Cut trees can substantially improve the scalability of these methods as a back-end.

\item \textit{Application 3}:
  As cut trees elegantly encode all the min-cuts (i.e., the min-cuts of ${n \choose 2}$ pairs) of the original graph in $O(n)$ size,
  we can design algorithms that extract interesting statistics about all the min-cuts from a cut tree
  in near-linear time without instantiating all the min-cuts.
  We discuss a few examples of algorithms
  that efficiently compute the \emph{connectivity distribution} and \emph{connectivity dendrogram}
  from cut trees in Section~\ref{sec:applications}.
\end{itemize}

\myparagraph{Scope}
We focus on real sparse graphs
such as social networks and web graphs,
and design an efficient algorithm tailored to these networks.
We do not claim that our algorithm is efficient for all kinds of graphs,
e.g., those arising from optimization problems.

\myparagraph{Organization}
The remainder of this paper is organized as follows.
In Section~\ref{sec:preliminaries},
we explain the basic notation and definitions used throughout this paper.
We present an overview of our cut-tree construction algorithm in Section~\ref{sec:algorithm}.
We discuss $s$-$t$ cut computation algorithms
tailored to real-world networks of interest in Section~\ref{sec:flow}.
In Sections~\ref{sec:packing} and \ref{sec:goal},
we propose two heuristics to efficiently find min-cuts
without running max-flow algorithms.
We explain how to select separation pairs in Section~\ref{sec:select}.
Section~\ref{sec:reduce} is devoted to the graph reduction rules.
We present our experimental results in Section~\ref{sec:experiments},
and discuss applications of cut trees to large-scale network analysis
in Section~\ref{sec:applications}.
We describe some previous work in this area in Section~\ref{sec:related_work}.
Finally, we conclude the paper in Section~\ref{sec:conclusions}.

\section{Preliminaries}
\label{sec:preliminaries}

\subsection{Notations and Definitions}

In this paper, we focus on networks that can be modeled as undirected graphs.
Let $G=(V,E)$ be an undirected graph.
We denote the degree of a vertex $v$ by $d(v)$.
For a vertex subset $S\subseteq V$ and a fresh vertex $s\not\in V$, we denote the graph obtained by contracting $S$ into $s$ as $G/(S\rightarrow s)$, i.e.,
the graph obtained by adding $s$, reconnecting all edges between $S$ and $V\setminus S$ to $s$, and removing $S$.
For a directed graph,
we denote the set of edges outgoing from vertex $v$ as $\delta^+(v)$ and the set of incoming edges as $\delta^-(v)$.

\subsection{Network Flows}

Let $G=(V,E,c)$ be a directed graph with an edge-capacity function $c:E\rightarrow\mathbb{R}_{\geq 0}$.
For two distinct vertices $s$ and $t$, a vertex subset $S\subseteq V$ is called an \emph{$s$-$t$ cut} if $S$ contains $s$ but does not contain $t$,
and its \emph{capacity} $c_G(S)$ is defined as the total capacity of the outgoing edges $\{uv\in E\mid u\in S,v\not\in S\}$.
An $s$-$t$ cut with the minimum capacity is called the \emph{minimum $s$-$t$ cut}.

A function $f:E\rightarrow\mathbb{R}_{\geq 0}$ is called an $s$-$t$ flow if it satisfies the following two conditions:
\begin{itemize}
  \item $0\leq f(e)\leq c(e)$ $(\forall e\in E)$, and
  \item $\sum_{e\in\delta^+(v)}f(e)=\sum_{e\in\delta^-(v)}f(e)$ $(\forall v\in V\setminus\{s,t\})$.
\end{itemize}
The \emph{value} of an $s$-$t$ flow $f$ is defined by $\mathrm{val}(f,s)=\sum_{e\in\delta^+(s)}f(e)-\sum_{e\in\delta^-(s)}f(e)$,
and an $s$-$t$ flow with the maximum value is called the \emph{maximum $s$-$t$ flow}.
The famous \emph{min-cut max-flow theorem} states that, for any graph, the capacity of the minimum $s$-$t$ cut is equal to the value of the maximum $s$-$t$ flow.

Let $f$ be an $s$-$t$ flow (which may not be maximum).
A \emph{residual graph} with respect to $f$ is a directed graph $G_f=(V, E_f)$ defined as
\[
E_f=\{e\in E\mid f(e)<c(e)\}\cup\{vu\mid uv\in E, f(uv)>0\}.
\]
An $s$-$t$ path in the residual graph $G_f$ is called an \emph{$f$-augmenting path}.
If there exists an $f$-augmenting path, we can obtain a greater flow $f'$.
A flow is maximum if and only if there are no $f$-augmenting paths.
For any maximum $s$-$t$ flow $f$, a set of vertices $S$ reachable from $s$ in $G_f$ becomes a minimum $s$-$t$ cut,
which is also \emph{minimal} among all the minimum $s$-$t$ cuts in the sense of set inclusion.

In this paper, we focus on undirected graphs with the unit-capacity function $c(e)=1$; however, most parts of the proposed algorithm can be applied to capacitated graphs.
In an undirected graph $G=(V,E)$, the cut and flow are defined by considering the bidirected graph $\bar{G}=(V,\bar{E})$ obtained from $G$
by replacing each undirected edge $uv$ with two directed edges $uv$ and $vu$.
In this setting, the capacity of the minimum $s$-$t$ cut is the number of edges that must be removed to separate $s$ and $t$ into different connected components.
Thus, this value is called the \emph{connectivity} between $s$ and $t$, which is denoted by $\lambda_G(s,t)$.

\subsection{Cut Trees}
For an undirected graph $G=(V,E)$, a tree $T=(V,E')$ on the same vertex set is called a \emph{cut tree} (or \emph{Gomory-Hu tree}) if it satisfies the following condition for any distinct vertices $s,t\in V$:
\[
\lambda_G(s,t)=\min_{e\in P_{st}}c_G(S_e),
\]
where $P_{st}$ is the unique path from $s$ to $t$ in the tree $T$, $S_e$ is the connected component of $T$ containing $s$ obtained by the removal of an edge $e$,
and $c_G(S_e)$ is the number of outgoing edges from $S_e$ in the graph $G$.
In other words, the condition states that at least one of the cuts $S_e$ induced by an edge $e$ on the path $P_{st}$ becomes the minimum $s$-$t$ cut.
For convenience, we will construct an edge-weighted cut tree $T=(V,E',c')$ satisfying $c'(e)=c_G(S_e)$ for all $e\in E'$.
Using such a tree, we can obtain the connectivity between two vertices $s$ and $t$ by simply computing the minimum $c'(e)$ value over the edges $e\in P_{st}$.

\begin{figure}[t!]
  \centering
  \includegraphics[width=1 \hsize]{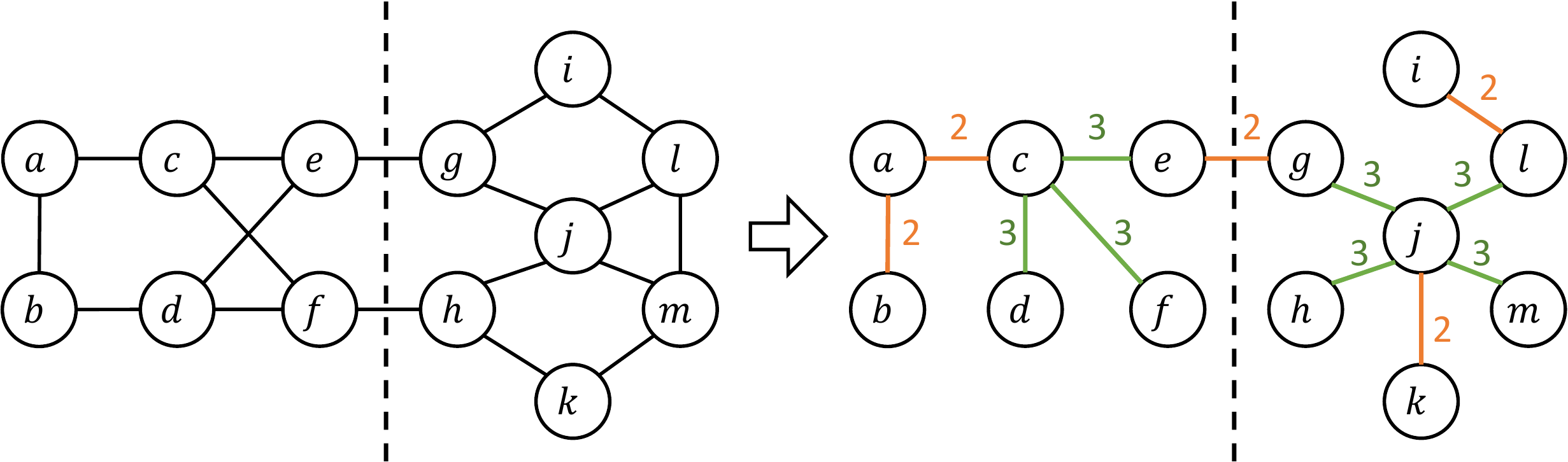}
  \caption{An example of the cut tree.}
  \label{fig:cut-tree-example}
\end{figure}

Figure~\ref{fig:cut-tree-example} shows an example of a graph (left) and its cut tree (right).
The orange-colored edges have a weight value of two and the green-colored edges have weight three.
Each edge in the tree induces a minimum cut in the original graph; e.g., the edge $eg$ in the cut tree induces a cut $\{a,b,c,d,e,f\}$ and this is the minimum $e$-$g$ cut in the original graph.
We can find the connectivity by identifying the minimum weight edge on the unique path; e.g., the unique path from $d$ to $j$ consists of edges $dc$, $ce$, $eg$, and $gj$,
and edge $eg$ has the minimum weight of two.
Thus, the connectivity between $d$ and $j$ is two.

\subsection{Basic Cut-Tree Construction Algorithm}

Algorithm~\ref{alg:basic} describes the basic algorithm developed by Gomory and Hu for constructing a cut tree~\cite{cut-tree/gomory-hu}.
In the algorithm, each vertex of tree $T$ corresponds to a subset of vertices $X\subseteq V$, which induces a partition of $V$.
To avoid confusion, we refer to the vertices of the tree as \emph{nodes}.
Initially, $T$ consists of only a single node $V$.
The algorithm iteratively picks a node $X$ of size at least two, and splits it into two smaller nodes (Procedure \textit{Separate}).
The details of this part are described later.
Finally, each node of $T$ corresponds to a single vertex, and we have obtained the cut tree.

For each node $X\subseteq V$, there is a corresponding graph $G_X$ on a vertex set $V_X\supseteq X$.
Vertices in $V_X\setminus X$ are called \emph{contracted}, and each contracted vertex $v$ corresponds to an edge $\phi(v)$ incident to the node $X$.
At first, the node $V$ corresponds to the original graph $G$, and there are no contracted vertices.

We now describe the details of Procedure \textit{Separate}.
When splitting a node $X$, the algorithm first picks an arbitrary pair $\{s,t\}$ from ${X \choose 2}$ and computes a minimum $s$-$t$ cut $S$.
Node $X$ is then split into two smaller nodes $X_s=X\cap S$ and $X_t=X\setminus S$.
These two nodes are connected by an edge $e$ whose capacity is equal to the capacity of the minimum $s$-$t$ cut.
Node $X_s$ corresponds to a graph $G_s$ obtained from $G_X$ by contracting the outside of the cut $S$ into a single vertex $t'$,
and node $X_t$ corresponds to a graph $G_t$ obtained by contracting the inside of the cut $S$ into a single vertex $s'$.
The two contracted nodes $s'$ and $t'$ are set to correspond to the newly introduced edge $e$.
Finally, the edges incident to node $X$ are reconnected as follows:
for each contracted vertex $v\in V_X\setminus X$ inside the cut $S$, the corresponding edge $\phi(v)$ is reconnected to $X_s$,
and for each other contracted vertex, the corresponding edge is reconnected to $X_t$.

\begin{figure*}[t!]
  \centering
  \includegraphics[width=1 \hsize]{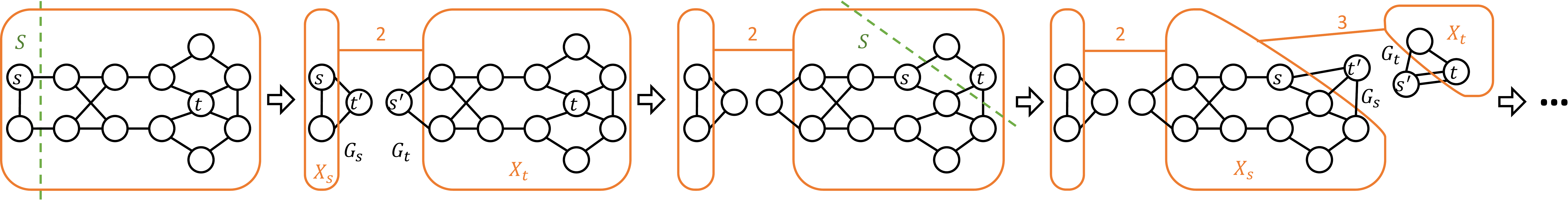}
  \caption{Example execution of the Gomory-Hu algorithm.}
  \label{fig:cut-tree-construction}
\end{figure*}

Figure~\ref{fig:cut-tree-construction} illustrates an example execution.
The green dotted lines show $s$-$t$ min-cuts, and the orange lines denote the tree and the corresponding sets.
At first, the tree consists of only a single node $V$.
A pair $\{s,t\}$ is selected and a min-cut is computed, as shown in the leftmost figure.
Node $V$ is then split into two sets and the graph is split into two contracted graphs, as shown in the second figure.
In the next step, another pair is selected from the right node, and the corresponding min-cut is computed, as in the third figure.
The right node is then split into two nodes and the edge incident to this node is reconnected to $X_s$, because the contracted node created in the first step is located inside the cut $S$.
This process is repeated until all the nodes become a singleton.

Note that the contracted graphs $G_s$ and $G_t$ are only created for efficiency --- we can correctly compute a cut tree
using the same graph $G_X$ instead of $G_s$ and $G_t$.
In this case, when reconnecting edges at lines~\ref{line:reconnect_loop}--\ref{line:reconnect_loop_end}, instead of using the contracted vertices $s'$ and $t'$, we can use arbitrary vertices in $S$ and $V_X\setminus S$, respectively\footnote{
One may think that a future min-cut $S'$ could cross the min-cut $S$, leaving us unable to determine which side of the cut the corresponding edge should reconnect with.
However, from the submodularity and posimodularity of the min-cut, using the minimal min-cut $S$ ensures that such a case never occurs. For details, see the paper by Gomory and Hu~\cite{cut-tree/gomory-hu}.}.

\begin{algorithm}[t!]
\small
\Procedure{\myfnsty{Construct-Basic}$(G = (V, E))$}{
  \nl \myfnsty{Init}$(G)$\;
  \nl \myfnsty{Separate-All}$()$\;
  \nl \Return{$T$}\;
}
\algrule
\Procedure{\myfnsty{Init}$(G = (V, E))$}{
  \nl $\mathcal{G} \gets \set{(G,V)}$; $T \gets (\{V\}, \emptyset, \emptyset)$; $\phi\gets\emptyset$\;
}
\algrule
\Procedure{\myfnsty{Separate-All}$()$}{
  \nl \While{there is $(G_X,X) \in \mathcal{G}$ such that $\abs{X} > 1$}{
    \nl $\set{s, t} \gets$ an arbitrary pair from ${X \choose 2}$\;\label{line:choose}
    \nl $\text{\myfnsty{Separate}}(G_X, X, s, t)$\;
  }
}
\algrule
\Procedure{\myfnsty{Separate}$(G_X=(V_X,E_X), X, s, t)$}{
  \nl $S \gets \text{\myfnsty{MinCut}}(G_X, s, t)$\;\label{line:mincut}
  \nl $X_s \gets X\cap S; X_t \gets X\setminus S$\;
  \nl Add new nodes $X_s$ and $X_t$ connected by an edge $e$ of capacity $\lambda_G(s,t)$ into $T$\;
  \nl $G_s \gets G_X/(V_X\setminus S\rightarrow t'); G_t \gets G_X/(S\rightarrow s')$\;\label{line:contract}
  \nl $\mathcal{G} \gets (\mathcal{G} \setminus \{(G_X,X)\}) \cup \set{(G_s,X_s), (G_t, X_t)}$\;
  \nl $\phi(s')\gets e; \phi(t')\gets e$\;
  \nl \For{each $v\in V_X\setminus X$}{\label{line:reconnect_loop}
    \nl \textbf{if} $v\in S$ \textbf{then} Reconnect $\phi(v)$ from $X$ to $X_s$\;
    \nl \textbf{else} Reconnect $\phi(v)$ from $X$ to $X_t$\;\label{line:reconnect_loop_end}
  }
  \nl Remove the node $X$ from $T$\;
}
\caption{Basic construction algorithm~\cite{cut-tree/gomory-hu}.}
\label{alg:basic}
\end{algorithm}

For practical efficiency, we apply the following three naive improvements to this basic algorithm.
First, when constructing contracted graphs $G_s$ and $G_t$ at line~\ref{line:contract}, instead of constructing them from scratch,
we reuse the original graph $G_X$ and convert it into graphs $G_s$ and $G_t$ by creating new vertices $\{s',t'\}$ and reconnecting the edges between $S$ and $V_X\setminus S$.
Because the graph $G_X$ is never used again, it does not need to be restored.

Second, when the size of the obtained cut is $1$, we do not construct the contracted graphs.
By renaming $s$ as $s'$, the graph $G_X$ is exactly the same as the contracted graph $G_t$.
Moreover, the other contracted graph $G_s$ is never used in the algorithm.
Thus, it is sufficient to set the vertex $s$ to correspond to the edge between nodes $\{s\}$ and $X\setminus\{s\}$.

Third, instead of traversing all the contracted vertices at line~\ref{line:reconnect_loop}, we only traverse the vertices inside the cut $S$ and reconnect the corresponding edges to the node $X_s$.
Then, instead of creating the node $X_t$ and reconnecting the remaining edges, we just rename the node $X$ as $X_t$.

Algorithm~\ref{alg:basic} contains two degrees of freedom: the first is how to select a pair $\{s,t\}$ at line~\ref{line:choose} and the second is how to compute the minimum $s$-$t$ cut at line~\ref{line:mincut}.
Although any selection strategy and any max-flow algorithm can correctly compute cut trees, the choice can have a significant effect on performance.

\section{Algorithm Overview}
\label{sec:algorithm}

In this section, we present an overview of our cut-tree construction algorithm.
The overall algorithm is described in Algorithm~\ref{alg:overview}.
To computing min-cuts efficiently, we
discuss the max-flow algorithms best suited to real networks of interest
and propose practical improvements in Section~\ref{sec:flow}.
In our algorithm, instead of finding individual min-cuts by computing a max-flow $|V|-1$ times, we detect multiple min-cuts at once
by \emph{tree packing} (line~\ref{line:our:packing}).
This technique is described in Section~\ref{sec:packing}.
The remaining graph is then separated by computing max-flows.
As we would still need to compute the max-flows for a huge number of vertex pairs, we do not compute each max flow from scratch,
but instead precompute some information to speed up the multiple computations (line~\ref{line:our:goal}).
This method, which is explained in Section~\ref{sec:goal}, is only applied to large components, and each separated component is processed by the basic method.
We explain how to select separation pairs in Section~\ref{sec:select}
(lines~\ref{line:our:separate1}--\ref{line:our:separate3}).
Finally, in Section~\ref{sec:reduce}, we explain the reduction rules applied at the beginning of the algorithm (lines~\ref{line:our:reduce1}--\ref{line:our:reduce2}) to reduce the size of the input graph.

\begin{algorithm}[t!]
\small
\Procedure{\myfnsty{Construct-Fast}$(G = (V, E))$}{
  \nl \myfnsty{Init}$(G)$\;
  \nl \tcp{Graph reduction.}
  \nl \myfnsty{Decompose-2-Connected-Components}$()$\;\label{line:our:reduce1}
  \nl \myfnsty{Contract-Degree-2-Vertices}$()$\;\label{line:our:reduce2}
  \nl \tcp{Heuristics to find a large portion of cuts.}
  \nl \myfnsty{Find-Cuts-by-Tree-Packing}$()$\;\label{line:our:packing}
  \nl \tcp{Accelerated max-flow computations by preprocessing.}
  \nl \myfnsty{Find-Cuts-by-Goal-Oriented-Search}$()$\;\label{line:our:goal}
  \nl \tcp{Separate the remaining components.}
  \nl \myfnsty{Separate-High-Degree-Pairs}$()$\;\label{line:our:separate1}
  \nl \myfnsty{Separate-Adjacent-Pairs}$()$\;
  \nl \myfnsty{Separate-All$()$}\;\label{line:our:separate3}
  \nl \Return{$T$}\;
}
\caption{Overall proposed construction algorithm.}
\label{alg:overview}
\end{algorithm}

\section{{\secit{s}}-{\secit{t}} Cut Computations}
\label{sec:flow}

We now discuss one of the most important building blocks: algorithms to compute minimum $s$-$t$ cuts.
Among the various methods for determining the max-flow, we focus on Dinitz's algorithm~\cite{1970:din}, which is described in Algorithm~\ref{alg:dinitz}.
This first constructs a shortest-$s$-$t$-path directed acyclic graph (DAG)\footnote{The shortest-$s$-$t$-path DAG of $G$ is a subgraph of $G$ consisting of only edges contained in some shortest path from $s$ to $t$.} $D$ in the residual graph $G_f$.
The flow is then augmented by identifying an $f$-augmenting path that uses only edges contained in $D$.
When no $f$-augmenting paths can be found (such a flow $f$ is called a \emph{blocking flow} with respect to $D$), the shortest-path DAG $D$ is updated.
This process is repeated until $t$ becomes unreachable from $s$ in $G_f$.
For uncapacitated networks, computing a blocking flow has linear time complexity in the size of the DAG $D$\footnote{
For a capacitated network, it is known to take $O(|V(D)||E(D)|)$ time; however, such a worst case rarely occurs in practice.}.
Thus, if the DAGs are small and can be found efficiently, the algorithm is fast.

\begin{algorithm}[t!]
\small
\Procedure{\myfnsty{Max-Flow}$(G = (V, E), s, t)$}{
  \nl $f\gets 0$\;
  \nl \While{there exists an $s$-$t$ path in $G_f$}{
  	\nl$D\gets\myfnsty{Construct-Shortest-Path-DAG}(G_f,s,t)$\;
  	\nl \While{there is an $f$-augmenting path $P$ in $D$}{
  		\nl Augment $f$ along $P$
  	}
  }
  \nl \Return{$f$}\;
}
\caption{Dinitz's max-flow algorithm~\cite{cut-tree/gomory-hu}.}
\label{alg:dinitz}
\end{algorithm}

Whereas Dinitz's original algorithm conducts a standard unidirectional breadth-first search (BFS),
we propose a bidirectional BFS to compute shortest-path DAGs, as this improves the practical efficiency on networks of interest.
The shortest-path computation itself is also important and has been the subject of considerable research.
These studies have found that, in real networks of interest, shortest-path DAGs are relatively small and can be efficiently constructed using bidirectional BFS~\cite{DBLP:journals/corr/BorassiN16}.
Although the graphs for which we need to construct shortest-path DAGs are actually residual graphs of the original real networks,
we found that the Dinitz's algorithm with bidirectional DAG construction works efficiently in preliminary experiments.

The bidirectional BFS for constructing a DAG is as follows.
We iteratively construct a set of vertices $S_d$ ($T_d$) that are located at distance $d$ from the vertex $s$ ($t$).
Initially, we set $S_0=\{s\}$, $T_0=\{t\}$, $d_s=0$, and $d_t=0$.
The following process is then repeated until the sets $S_{d_s}$ and $T_{d_t}$ intersect;
if the number of edges incident to $S_{d_s}$ is smaller than that of $T_{d_t}$, we compute $S_{d_s+1}$ by traversing the edges incident to $S_{d_s}$ and increase $d_s$ by one;
otherwise, we do the same for $T_{d_t}$.
This procedure gives the distances of vertices contained in balls of radius $d_s$ and $d_t$ from $s$ and $t$, respectively.
Finally, by running a reverse-BFS from $S_{d_s}\cap T_{d_t}$ using only the edges from $S_{i+1}$ to $S_i$, or from $T_{i+1}$ to $T_i$, we can construct the shortest-$s$-$t$-path DAG.
In our implementation, we do not explicitly construct the DAG, but only compute the distances; when computing augmenting paths, we only use edges from $S_i$ to $S_{i+1}$ or from $T_{i+1}$ to $T_i$.

\myparagraph{Other practical max-flow algorithms}
The push-relabel method~\cite{DBLP:journals/jacm/GoldbergT88} is often considered as the best method in practice.
However, this approach does not involve a shortest-path computation, and it would appear to be difficult to make it bidirectional.
Thus, it does not allow the structure of real networks to be exploited.
For segmentation tasks in computer vision, Goldberg~et~al.~\cite{DBLP:conf/esa/GoldbergHKTW11} proposed a practical max-flow algorithm called IBFS
that uses a bidirectional search.
For the networks appeared in the segmentation tasks, the initial DAG already contains all the vertices in the graph.
Thus, simply constructing DAGs in a bidirectional manner does not enhance the computation speed.
Their approach expresses the DAG as two shortest-path trees that are dynamically updated after each augmentation.
However, in the networks we are interested in, the initial DAG is very small and grows through such augmentations; hence, this dynamic update approach would not lead to a significant speed-up.

\section{Greedy Tree Packing}\label{sec:packing}
Although we have designed a fast max-flow algorithm, computing $|V|-1$ max-flows to construct a cut tree remains very time consuming; e.g., even if a single max-flow can be computed in only $10$ ms, it would take $100,000$ s to compute all necessary max-flows in a graph with 10 million vertices.
To develop a much faster cut-tree algorithm, we must find min-cuts without relying on max-flow algorithms.
In this section, we propose a \emph{greedy tree packing} heuristic that identifies min-cuts between multiple pairs at once without using a max-flow algorithm.

For a directed graph and a vertex $r$, a subgraph is called an $r$-tree if (i) its underlying undirected graph is a connected tree (which may not be a spanning tree), (ii) $r$ has no incoming edges, and (iii) all other vertices in the tree have in-degrees of exactly one.
For an undirected graph $G=(V,E)$, a set $\mathcal{T}$ of edge-disjoint $r$-trees of the bidirected graph
$\bar{G}=(V,\bar{E})$\footnote{$\mathcal{T}$ can contain both directed edges $uv$ and $vu$ corresponding to a single undirected edge.}
is called an \emph{$r$-tree packing} of $G$.
The vertex $r$ is called the \emph{root} of the tree packing.
The following relationship between an $r$-tree packing and the edge-connectivity was derived by Bang{-}Jensen~et~al.~\cite{DBLP:journals/siamdm/Bang-JensenFJ95}.
For an undirected graph $G=(V,E)$ and its vertices $r,v\in V$, if there exist $k$ edge-disjoint paths from $r$ to $v$, the connectivity between $r$ and $v$ is at least $k$.
Such edge-disjoint paths can be composed into a single set of edge-disjoint $r$-trees for all $v\in V\setminus\{r\}$:
if there exists an $r$-tree packing $\mathcal{T}$ of an undirected graph $G=(V,E)$, the connectivity between $r$ and $v$ is at least the number of $r$-trees in $\mathcal{T}$ containing $v$.
Moreover, they showed that the converse also holds: there exists an $r$-tree packing $\mathcal{T}$ such that, for any vertex $v\in V\setminus\{r\}$, exactly $\lambda_G(r,v)$ $r$-trees in $\mathcal{T}$ contain $v$.

In our algorithm, we greedily construct an $r$-tree packing $\mathcal{T}$.
The details of this greedy algorithm are explained later.
The constructed tree packing $\mathcal{T}$ may not contain each vertex $v$ $\lambda_G(r,v)$ times; however,
if a vertex $v$ appears exactly $d(v)$ times in $\mathcal{T}$, we can confirm that the connectivity between $v$ and $r$ is exactly $d(v)$, and thus the cut $\{v\}$ is the minimum $v$-$r$ cut.
After constructing an $r$-tree packing, we can detect all such vertices $v$ and separate pairs $\{v,r\}$ in linear time.
We apply this strategy $\alpha$ times by selecting each of the top $\alpha$ degree vertices as the root $r$.
The effects of the parameter $\alpha$ are discussed in Section~\ref{sec:experiments}.

Our greedy packing algorithm proceeds as follows.
Starting from the bidirected graph $\bar{G}$, we iteratively construct an $r$-tree and remove its edges from $\bar{G}$.
Basically, we do not want to create dead ends; if we remove all outgoing edges from a vertex $v$, it will become a leaf in the subsequent tree construction.
If we use the BFS to construct an $r$-tree, the first tree removes all the outgoing edges of $r$, and we cannot construct a second tree.
Thus the BFS should not be used.
In order to avoid creating such dead ends, we use a depth-first search (DFS).

Additionally, we restrict the out-degree of each vertex in an $r$-tree to being at most $\beta$.
If the current visiting vertex in the DFS is $v$ and we have already used $\beta$ edges from $\delta^+(v)$ in the current tree, we immediately backtrack from vertex $v$ to its parent without using the remaining edges in $\delta^+(v)$.
A larger value of $\beta$ would find a larger tree, but may create more dead ends.
We discuss the trade-off effects of different $\beta$ values in Section~\ref{sec:experiments}.

\begin{figure}[t!]
  \centering
  \includegraphics[width=0.65 \hsize]{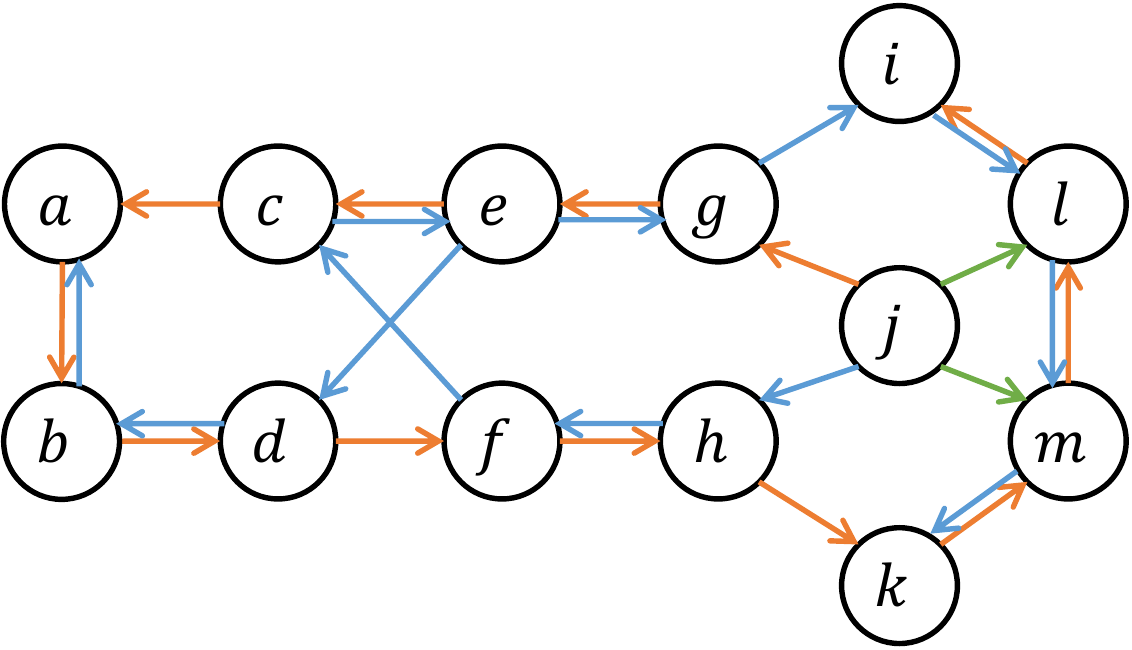}
  \caption{An example of the tree-packing.}
  \label{fig:tree-packing}
\end{figure}

Figure~\ref{fig:tree-packing} illustrates a tree-packing constructed by the greedy DFS.
Here, the highest degree vertex $j$ is chosen as the root, and the three trees are colored orange, blue, and green.
The degree-two vertices $\{a,b,i,k\}$ appear twice in the tree-packing and a subset of degree-three vertices $\{l,m\}$ appear three times.
Thus, for each of these vertices, we can immediately obtain a min-cut.
Because the remaining degree-three vertices $\{c,d,e,f,g,h\}$ appear only twice, they cannot be separated by this tree-packing, and are processed by other tree-packings or different methods.

\section{Goal-Oriented Search}\label{sec:goal}
To construct a cut tree, we must compute the maximum $s$-$t$ flows multiple times.
Instead of computing each max-flow from scratch, we propose to precompute some information and accelerate these multiple computations
for certain kinds of vertex pairs.

In the Gomory-Hu algorithm, we are free to choose the separation pairs.
To make the necessary precomputation possible, instead of selecting an arbitrary pair,
we first fix a sink $t\in X$.
When separating a set $X$ containing $t$, we then always choose a pair $\{s,t\}$ for some vertex $s\in X$.
Using this selection strategy, we can compute the initial shortest-$s$-$t$-path DAG used in Dinitz's algorithm more efficiently than using the bidirectional BFS.
First, we precompute a shortest path tree from the vertex $t$.
When processing a pair $\{s,t\}$, we construct a shortest-$s$-$t$-path DAG using DFS from vertex $s$ and only using edges $uv$
for which the distance from $u$ to $t$ is exactly the distance from $v$ to $t$ plus one, i.e., a shortest path from $u$ to $t$ passes vertex $v$.
To avoid updating the shortest-path tree after the contraction, we only create the contracted graph $G_s$.
Instead of creating $G_t$, we reuse $G_X$.
As explained in Section~\ref{sec:preliminaries}, such a modification does not affect the correctness of the algorithm.

As this construction only visits vertices contained in the constructed DAG, it is much faster than the bidirectional BFS.
Computing the blocking flow has linear time complexity with respect to the size of the DAG.
Thus, if the first blocking flow becomes the maximum flow, this strategy leads to a significant speed-up.
However, if the first blocking flow is not the maximum, we need to update the shortest-$s$-$t$-path DAG.
As this second DAG computation uses the residual graph rather than the original graph, we cannot use the precomputed shortest-path tree.

To avoid time-consuming DAG updates, we search for augmenting paths that use edges not on the DAG.
In addition to the edges $uv$ for which the distance from $u$ to $t$ is exactly the distance from $v$ to $t$ plus one, we also allow the use of \emph{detour} edges
$uv$ for which the distance from $u$ to $t$ is equal to the distance from $v$ to $t$.
The resulting graph might not be a DAG and could contain loops.
When searching for an augmenting path from $s$ to $t$, we allow the use of at most $\gamma$ detour edges, where $\gamma$ is a parameter.
This can be done by extending each vertex $v$ to a set of vertices $\{v_0, \ldots, v_\gamma\}$, adding an edge $u_iv_i$ for each edge $uv$ contained in the original DAG,
and adding an edge $u_iv_{i+1}$ for each detour edge $uv$.
A larger value of $\gamma$ will produce more augmenting paths and increase the likelihood of finding a maximum flow, but will have a higher computation time.
We discuss the trade-off effects of $\gamma$ in Section~\ref{sec:experiments}.

As the networks of interest tend to have unbalanced cuts that separate a small set of vertices from the remaining large set of vertices,
we only apply this strategy against the initial set $V$ and the highest degree vertex $t$.

\section{Selecting Separation Pairs}\label{sec:select}
In this section, we discuss how to select the next separation pair
among the remaining vertex sets after the goal-oriented search.
In general, there are two choices: select a pair with a \emph{balanced} min-cut to make the graphs obtained by the contraction smaller, or select a pair whose min-cut is easy to compute.
Here, the term balanced means that both $|S|$ (the inside of the cut) and $|V_X\setminus S|$ (the outside of the cut) are large.

\subsection{High Degree Pairs}
Goldberg and Tsioutsiouliklis~\cite{cut-tree/goldberg2001} developed heuristics to find such balanced min-cuts.
However, the networks of interest to us do not seem to have well-balanced min-cuts.
For example, it would be surprising if a social graph of 2 million vertices could be split into two components of 1 million vertices just by removing 100,000 edges.
Thus, it is not important to make the cut balanced, and it is better to focus on the pairs whose min-cut can be easily computed.
In our algorithm, we attempt to make the graphs smaller by finding somewhat balanced cuts.
Hence, we try to split the top-$k$ degree vertices before moving to the second selection strategy.
In this study, we use $k=10$.
We split large-degree vertices because the size of the min-cut is at most the size of the trivial cut $\{s\}$, which is equal to the degree of $s$,
and therefore small-degree vertices are less to have balanced cuts than high-degree vertices.

\subsection{Adjacent Pairs}
If the distance between $s$ and $t$ is $d$, the bidirectional BFS visits vertices contained in balls whose radius is approximately $d/2$ from $s$ and $t$.
Therefore, the smaller the distance, the faster the bidirectional BFS procedure, and a cut between nearby vertices would be easy to find using the bidirectional form of Dinitz's algorithm.
In our algorithm, we choose a pair $\{s,t\}\subseteq X$ such that $s$ and $t$ are adjacent in $G_X$.
If there are no such pairs, we choose an arbitrary pair from the remaining vertices.
Note that such a case can actually occur: consider a graph $G=(\{s,t,a,b,c\}, \{sa,sb,sc,at,bt,ct\})$; after separating pairs $\{s,a\}$, $\{s,b\}$, and $\{s,c\}$, we need to separate the non-adjacent pair $\{s,t\}$.

\section{Graph Reductions}\label{sec:reduce}
To reduce the size of the input graph before applying the algorithm, we use the following two strategies.

\subsection{Decomposing 2-Connected Components}
Let us assume that the input graph is connected; otherwise, we can construct cut trees separately for each connected component.
We can compute all the cuts of size 1, called \emph{bridges}, in linear time~\cite{tarjan1974note}.
For any pair that is not separated by bridges, the max-flow does not pass the bridges.
Thus, we can simply remove all the bridges and deal with each 2-connected component separately.

This reduction not only reduces the graph size, but also has a positive effect on the greedy tree packing heuristic described in Section~\ref{sec:packing}.
If a vertex $v$ has a neighbor $u$ of degree 1, $u$ can only become a leaf of a tree, and no $r$-trees can use the edge from $u$ to $v$.
Thus, such a vertex $v$ cannot appear $d(v)$ times in the constructed disjoint $r$-trees.
After applying this reduction, all vertices of degree 1 are removed, and therefore more min-cuts can be found by greedy tree packing.

\subsection{Contracting Degree-2 Vertices}
If there is a vertex $v$ of degree 2, the connectivity between $v$ and any other vertex in the same 2-connected component is exactly 2.
For any other vertices $s$ and $t$, if an $s$-$t$ flow uses one of the edges incident to $v$, it must use the other one.
Thus, we can replace vertex $v$ and its incident edges with an edge connecting the neighbors of $v$.

{
  \tabcolsep=2.1mm
\begin{table*}
\caption{Dataset information and cut-tree construction time (s). DNF denotes that the algorithm did not finish within 10 h.}
\label{tbl:construction-time}
\centering
\begin{tabular}{lrr|rrrrrr|rr}
  \toprule
  \multicolumn{3}{c|}{\textbf{Dataset}} & \multicolumn{6}{c|}{\textbf{Proposed Algorithms}}
  & \multicolumn{2}{c}{\textbf{Baselines}}\\
  Name &           $\abs{V}$ &             $\abs{E}$ & A0 & A1 &      A2 &    A3 &    A4 &    A5 & \textsc{GHg}~\cite{cut-tree/goldberg2001} &  Lemon~\cite{lemon} \\
\midrule
ca-GrQc          &       $5,242$ &         $14,484$ &     $3.0$ &         $0.1$ &     $0.1$ &     $0.1$ &     $0.1$ &    $<0.1$ &        $4.0$ &     $2.2$ \\
ca-CondMat       &      $23,133$ &         $93,439$ &   $110.1$ &         $1.4$ &     $1.2$ &     $0.8$ &     $0.7$ &     $0.2$ &      $256.7$ &    $75.7$ \\
soc-Epinions1    &      $75,879$ &        $405,740$ &   $977.2$ &         $7.0$ &     $4.2$ &     $5.0$ &     $3.8$ &     $0.7$ &     $2444.1$ &  $1113.5$ \\
com-DBLP         &     $317,080$ &      $1,049,866$ &       DNF &        $72.7$ &    $62.9$ &    $28.9$ &    $24.2$ &     $9.9$ &    $28212.1$ & $22098.0$ \\
com-Youtube      &   $1,134,890$ &      $2,987,624$ &       DNF &       $307.0$ &   $164.0$ &   $134.7$ &    $76.3$ &     $9.6$ &          DNF &       DNF \\
web-Google       &     $875,713$ &      $4,322,051$ &       DNF &       $365.1$ &   $294.7$ &    $62.6$ &    $43.9$ &    $50.2$ &          DNF &       DNF \\
web-BerkStan     &     $685,230$ &      $6,649,470$ &       DNF &       $569.9$ &   $530.8$ &   $327.1$ &   $102.7$ &    $40.3$ &          DNF &       DNF \\
soc-Pokec        &   $1,632,803$ &     $22,301,964$ &       DNF &      $3501.7$ &  $2996.9$ &  $2942.8$ &  $2216.7$ &    $71.8$ &          DNF &       DNF \\
soc-LiveJournal1 &   $4,847,571$ &     $42,851,237$ &       DNF &           DNF &       DNF &  $9543.3$ &  $6043.3$ &  $3178.2$ &          DNF &       DNF \\
hollywood-2011   &   $2,180,759$ &    $114,492,816$ &       DNF &     $22681.3$ & $20622.3$ & $11649.8$ & $10392.1$ &   $533.8$ &          DNF &       DNF \\
com-Orkut        &   $3,072,441$ &    $117,185,083$ &       DNF &           DNF &       DNF & $23376.3$ & $16511.2$ &   $679.0$ &          DNF &       DNF \\
indochina-2004   &   $7,414,866$ &    $150,984,819$ &       DNF &           DNF &       DNF &  $7254.4$ &  $1880.2$ &  $1169.3$ &          DNF &       DNF \\
arabic-2005      &  $22,744,080$ &    $553,903,073$ &       DNF &           DNF &       DNF &       DNF & $20961.2$ & $17836.6$ &          DNF &       DNF \\
it-2004          &  $41,291,594$ &  $1,027,474,947$ &       DNF &           DNF &       DNF &       DNF &       DNF & $28792.3$ &          DNF &       DNF \\
twitter-2010     &  $41,652,230$ &  $1,202,513,046$ &       DNF &           DNF &       DNF &       DNF &       DNF & $15323.1$ &          DNF &       DNF \\
\bottomrule
\end{tabular}
\end{table*}
}

{
  \tabcolsep=3.0mm
\begin{table}
\caption{Data size and average query time.}
\label{tbl:size_and_query}
\centering
\begin{tabular}{l|rr|r}
\toprule
\multicolumn{1}{c|}{\textbf{Dataset}} &
\multicolumn{1}{c}{\textbf{Graph size}} &
\multicolumn{1}{c|}{\textbf{Cut-tree size}} &
\multicolumn{1}{c}{\textbf{Query time}} \\
\midrule
ca-GrQc          &       56KB &                40KB &  0.069 $\mu$s \\
ca-CondMat       &      364KB &               180KB &  0.073 $\mu$s \\
soc-Epinions1    &      1.5MB &               0.6MB &  0.101 $\mu$s \\
com-DBLP         &      4.0MB &               2.4MB &  0.119 $\mu$s \\
com-Youtube      &     11.4MB &               8.7MB &  0.142 $\mu$s \\
web-Google       &     16.5MB &               6.7MB &  0.181 $\mu$s \\
web-BerkStan     &     25.4MB &               5.2MB &  0.177 $\mu$s \\
soc-Pokec        &     85.1MB &              12.5MB &  0.168 $\mu$s \\
soc-LiveJournal1 &    163.5MB &              37.0MB &  0.337 $\mu$s \\
hollywood-2011   &    436.8MB &              16.6MB &  0.139 $\mu$s \\
com-Orkut        &    447.0MB &              23.4MB &  0.137 $\mu$s \\
indochina-2004   &    576.0MB &              56.6MB &  0.405 $\mu$s \\
arabic-2005      &      2.1GB &               0.2GB &  0.469 $\mu$s \\
it-2004          &      3.8GB &               0.3GB &  0.369 $\mu$s \\
twitter-2010     &      4.5GB &               0.3GB &  0.294 $\mu$s \\
\bottomrule
\end{tabular}
\end{table}
}

\begin{figure*}[t!]
\centering
\subfloat[Number of tree packings ($\alpha$)]{
    \includegraphics[width=0.33 \hsize]{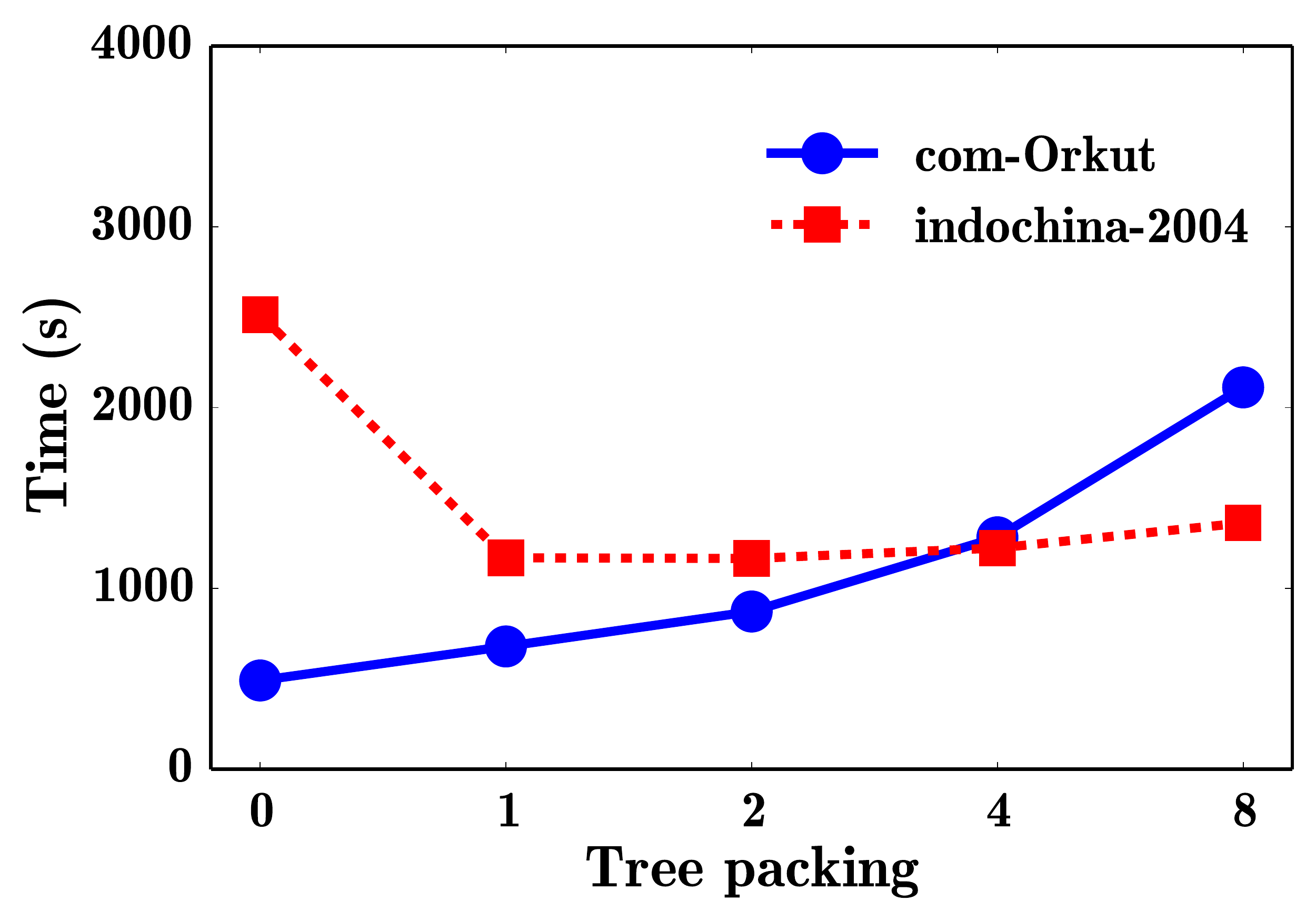}
    \label{fig:number_of_tree_packing}
}
\subfloat[Breadth limit of tree packing ($\beta$)]{
    \includegraphics[width=0.33 \hsize]{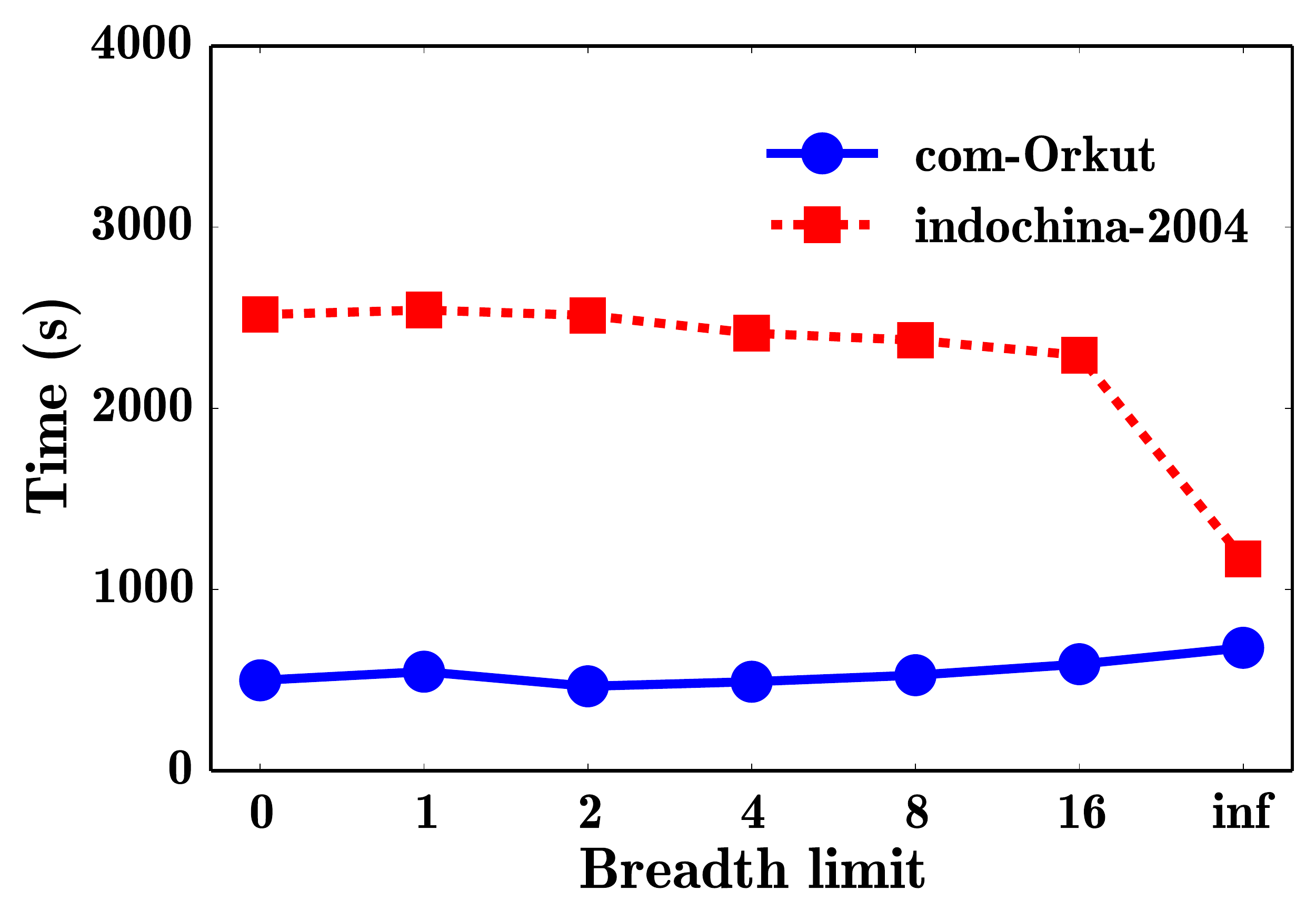}
    \label{fig:breadth_limit}
}
\subfloat[Search relaxation ($\gamma$)]{
    \includegraphics[width=0.33 \hsize]{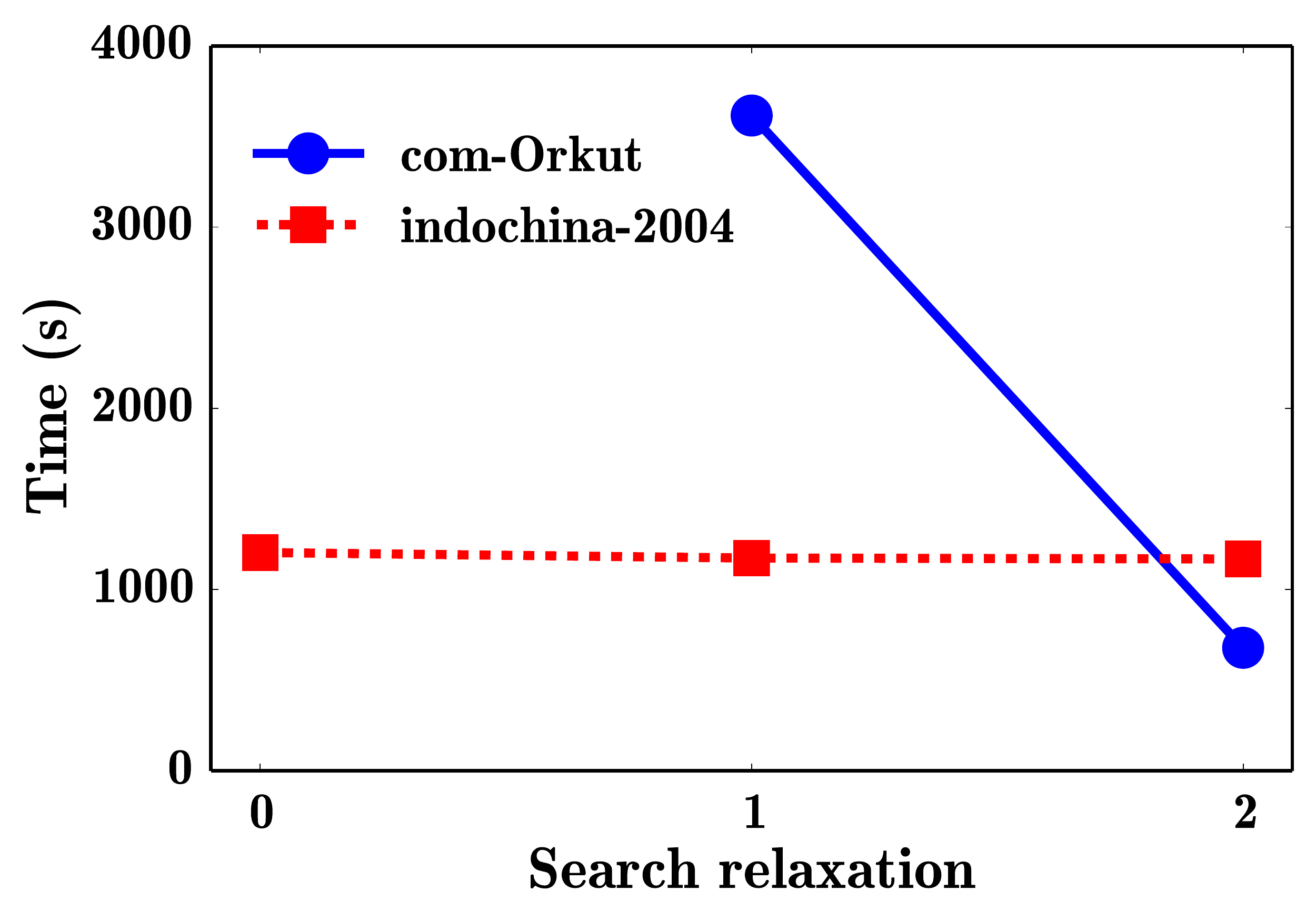}
    \label{fig:search_relaxation}
}
\caption{Runtime for various performance parameters.}
\end{figure*}

\section{Experimental Evaluation}
\label{sec:experiments}

\subsection{Setup}
\myparagraphF{Environment}
We conducted experiments on a Linux machine with an Intel Xeon X5650 processor ($2.67$ GHz) and $96$ GB main memory.
All algorithms were implemented in C++ and compiled using gcc 4.8.4 with the \texttt{-O3} option.

\myparagraph{Algorithms}
We compared the proposed method with two state-of-the-art cut-tree construction algorithms.
\emph{(1)} \textsc{GHg}~\cite{cut-tree/goldberg2001},
which combines the Gomory--Hu algorithm with balanced min-cut heuristics and the Hao--Orlin algorithm~\cite{hao1994faster}.
\emph{(2)} Lemon~\cite{lemon},
which is a highly tuned implementation of combinatorial optimization algorithms.
For the proposed algorithm,
unless otherwise stated,
we set the number of tree packings $\alpha = 1$, the breadth limit parameter $\beta = \infty$ (i.e., no limit), and
the search relaxation parameter $\gamma = 2$,
which we recommend as a robust setting.

\myparagraph{Datasets}
We used real-world social and web graphs that are publicly available from
the Stanford Large Network Dataset Collection~\cite{dataset/snap}
and Laboratory for Web Algorithms~\cite{dataset/webgraph1, dataset/webgraph2}.
Table~\ref{tbl:construction-time} summarizes the number of vertices and edges in these datasets.
The web-Google, web-BerkStan, indochina-2004, arabic-2005, and it-2004 datasets are web graphs;
the others are social graphs.

\subsection{Construction}
Our main focus is on reducing the cut-tree construction time.
We compared the following six versions of the proposed algorithm.
\underline{\emph{A0}} is the plain Gomory--Hu algorithm with a standard implementation of Dinitz's max-flow algorithm.
\underline{\emph{A1}} is another implementation of the Gomory--Hu algorithm using the bidirectional blocking flow algorithm introduced in Section~\ref{sec:flow}.
\underline{\emph{A2}} uses the same bidirectional blocking flow algorithm,
but employs graph reduction techniques such as 2-connected component decomposition and degree-2 vertex contraction,
and applies the high-degree pair separation strategy.
In addition to the above, \underline{\emph{A3}} employs the adjacent pair separation strategy,
and \underline{\emph{A4}} also uses greedy tree packing.
\underline{\emph{A5}} further conducts the goal-oriented search.
A5 is the overall proposed algorithm using the whole set of new techniques,
and is thus equivalent to Algorithm~\ref{alg:overview}.

Table~\ref{tbl:construction-time} lists the construction times achieved by each algorithm.
We wish to emphasize that
the A5 algorithm, which includes all of the proposed techniques, successfully constructed
cut trees for billion-scale web and social graphs (it-2004 and twitter-2010)
in 8 h and 4 h, respectively.
The baseline methods, \textsc{GHg} and Lemon, took several hours for
a million-scale social graph (com-DBLP),
and failed to construct complete cut trees for larger networks within the time limit of 10 h.
Therefore, the proposed method improves the scalability of cut-tree construction
by several orders of magnitude.

The results from different versions of the proposed method
show that more datasets were successfully processed within the time limit
as more of the new techniques were employed (i.e., from A0 to A5),
and the time required to treat each dataset consistently decreased.
These results indicate that
most of the proposed techniques are effective and essential for scalable cut-tree construction.

\subsection{Data Size and Query Time}
To confirm the practicality of cut trees,
we briefly discuss their data size and query times.
Note that these metrics are independent of the construction algorithm
(except the ways to break arbitrariness).

The data sizes of the resulting cut trees are listed in Table~\ref{tbl:size_and_query},
together with those of the original graphs.
It is clear that the cut trees are much smaller than the graphs.
This is as expected, as the graph and cut tree have sizes $\Theta(\abs{E})$
and $\Theta(\abs{V})$, respectively.

Table~\ref{tbl:size_and_query} also gives
the average query time for computing the $s$-$t$ cut size from the cut trees
for $10^7$ random pairs of vertices.
Using a naive query algorithm that simply ascends the trees from both ends,
the average query time is very small at less than 1 $\mu$s.
This is because the cut trees for these real graphs tend to be very shallow.

\subsection{Parameter Analysis}
Finally, we discuss the effect of different parameter values.
In these experiments,
we used a social network dataset, com-Orkut, and a web graph dataset, indochina-2004.
The trend for these two networks
can generally be observed in other social and web graphs.

\myparagraph{Number of tree packings}
Figure~\ref{fig:number_of_tree_packing}
illustrates the construction time for various values of $\alpha$, which is the number of tree packings.
From the results for $\alpha \geq 1$, we see that
applying tree packing multiple times is not beneficial.
The results for $\alpha=0$ and $1$
indicate that tree packing is effective for indochina-2004, but is not effective for com-Orkut.
The same trend can be observed for other web and social graphs.

\myparagraph{Breadth limit of tree packing}
Figure~\ref{fig:breadth_limit} shows the construction time for different value of $\beta$, which is the breadth limit of searches during tree packing.
For indochina-2004, setting $\beta=\infty$ (no breadth limit) results in construction that are approximately twice as fast as for other settings.
This is why we generally recommend $\beta=\infty$ for robustness.
In contrast, for com-Orkut, enabling the breadth limit
accelerates the algorithm up to 1.5 times.
In general,
$\beta$ should be set to a moderate constant   when handling social networks.

\myparagraph{Search relaxation}
Figure~\ref{fig:search_relaxation} shows the construction time for various $\gamma$,
which is the maximum number of detour steps allowed during goal-oriented searches.
It can be observed that a small positive
value of $\gamma$ drastically reduces the runtime for com-Orkut.
Indeed, with $\gamma=0$, the algorithm did not finish within the time limit.
In contrast, changes in the value of $\gamma$ had relatively little effect with the indochina-2004 dataset.
This is because indochina-2004 was separated earlier by balanced cuts.
In general, web graphs tend to have more balanced cuts than social graphs,
and search relaxation is more effective for social graphs.

\section{Applications}
\label{sec:applications}
We now discuss some applications of cut trees
to demonstrate the utility of the proposed algorithm.

\subsection{Connectivity Distribution}
The common structural properties of real networks are of interest to the data mining community,
although they have not yet been comprehensively studied.
We believe that the \emph{connectivity distribution} represents a new tool for the structural analysis of networks,
and have designed an efficient algorithm using cut trees.

We define the connectivity distribution as the distribution of connectivity between every pair of vertices.
More specifically, the connectivity distribution of a graph $G$ is $\set{f_G(k)}_k$,
where $f_G(k)$ denotes the number of vertex pairs whose connectivity is $k$.
Note that $\sum_k f_G(k) = {n \choose 2}$.

As the total number of pairs is quadratic,
it is reasonable to assume that its computation will require quadratic time.
However, we propose an algorithm
that exactly computes the connectivity distribution in $O(n \log n)$ time for a given cut tree.
The procedure is described in Algorithm~\ref{alg:connectivity_distribution}.
For each edge in the cut tree, the underlying idea of the algorithm is to count the number of pairs
with that corresponding minimum cut.

\begin{algorithm}[t!]
\small
\Procedure{\myfnsty{Connectivity-Distribution}$(T = (V_T, E_T, c_T))$}{
  \nl $C = \set{c_T(e) \mid e \in E_T}$; $\mathcal{V} \gets \set{\set{v} \mid v \in V_T}$;  $f \gets \emptyset$\;
  \nl \For{$x \in C$ in decreasing order}{
    \nl $f(x) \gets 0$\;
    \nl \For{$e \in E_T$ such that $c(e) = x$}{
      \nl $S, T \gets $ sets in $\mathcal{V}$ to which the ends of $e$ belong\;
      \nl $f(x) \gets f(x) + \abs{S} \times \abs{T}$\;
      \nl $\mathcal{V} \gets (\mathcal{V} \setminus \set{S, T}) \cup (S \cup T)$\;
    }
  }
  \nl \Return{$f$}\;
}
\caption{Computing the connectivity distribution.}
\label{alg:connectivity_distribution}
\end{algorithm}

\begin{figure}[t!]
  \centering
  \includegraphics[width=0.8 \hsize]{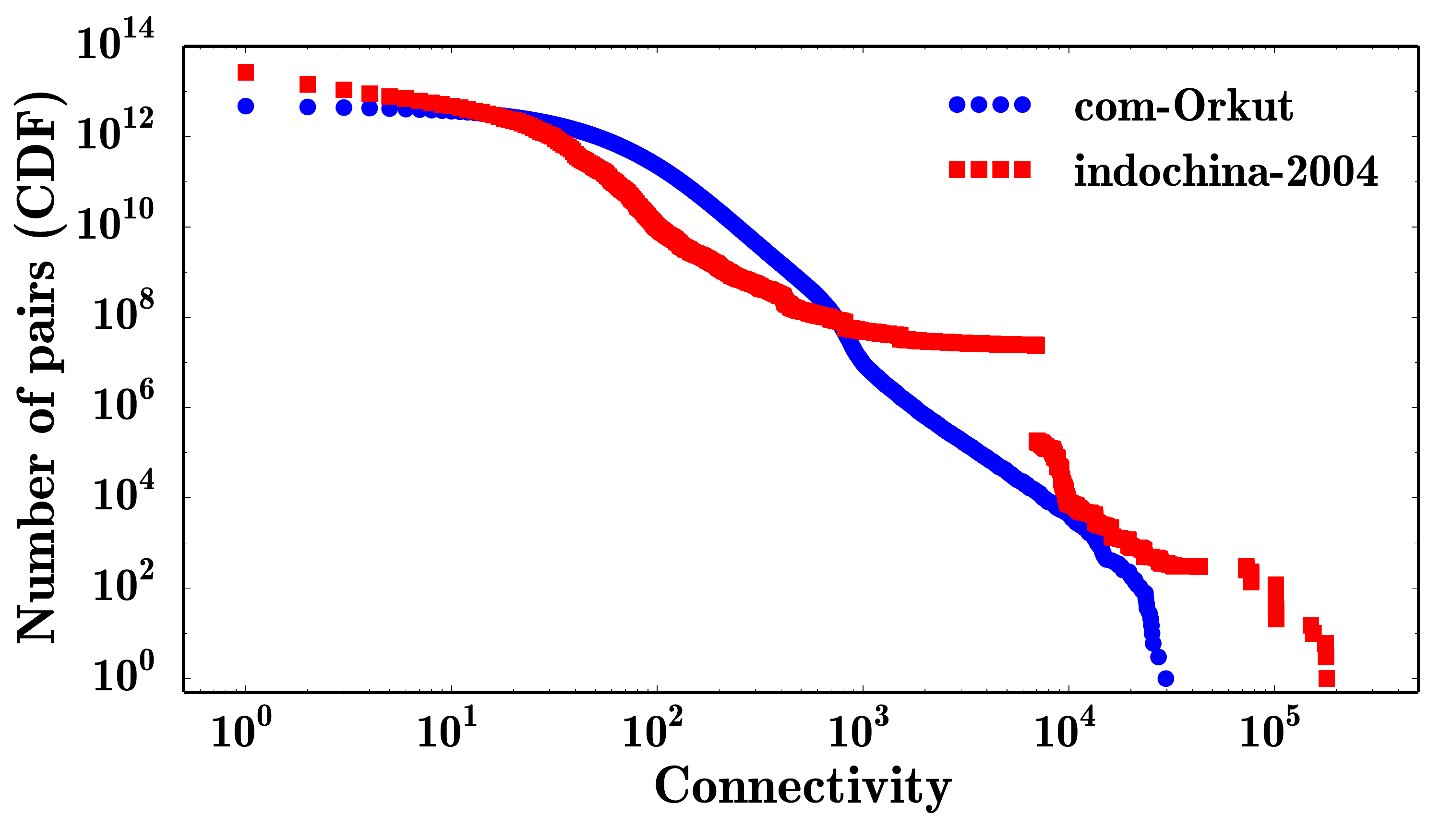}
  \caption{Examples of connectivity distribution.}
  \label{fig:connectivity_distribution}
\end{figure}

In experiments, this algorithm required only 0.06 s and 0.12 s
for cut trees of the com-Orkut and indochina-2004 datasets, respectively (excluding the time taken to construct the cut trees),
as illustrated in Figure~\ref{fig:connectivity_distribution}.
Interestingly, the connectivity seems to follow a power law, similar to the degree distributions.
However, deeper analysis of these distributions is beyond the scope of this paper.
We emphasize that our algorithms enable the connectivity distributions of large-scale networks to be studied for the first time.

\subsection{Connectivity Dendrogram}
\begin{figure}[t!]
\centering
\subfloat[A graph]{
\label{fig:dendrogram:graph}
\includegraphics[width=.49 \hsize]{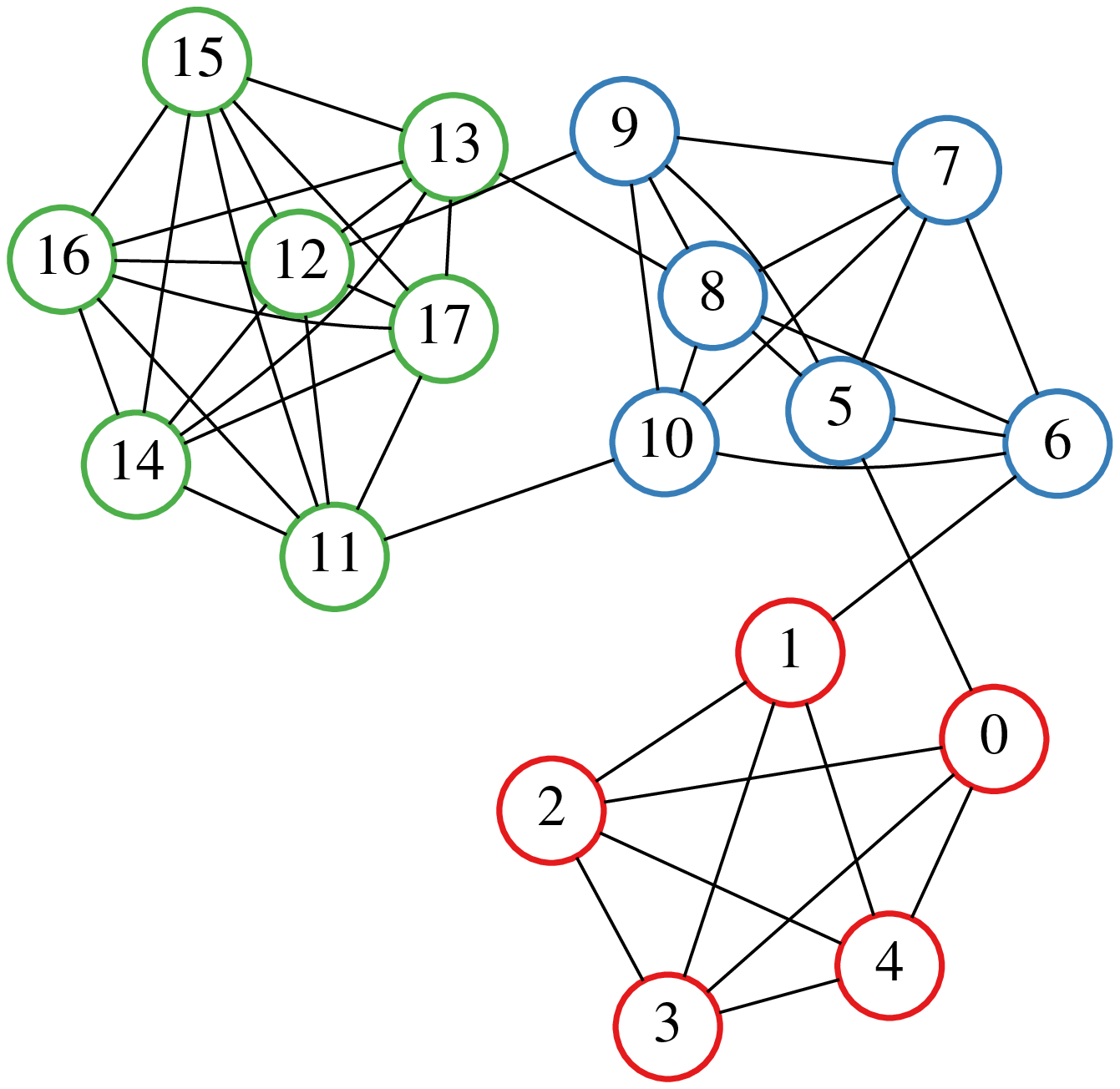}
}
\subfloat[The dendrogram]{
\label{fig:dendrogram:dendrogram}
\includegraphics[width=.49 \hsize]{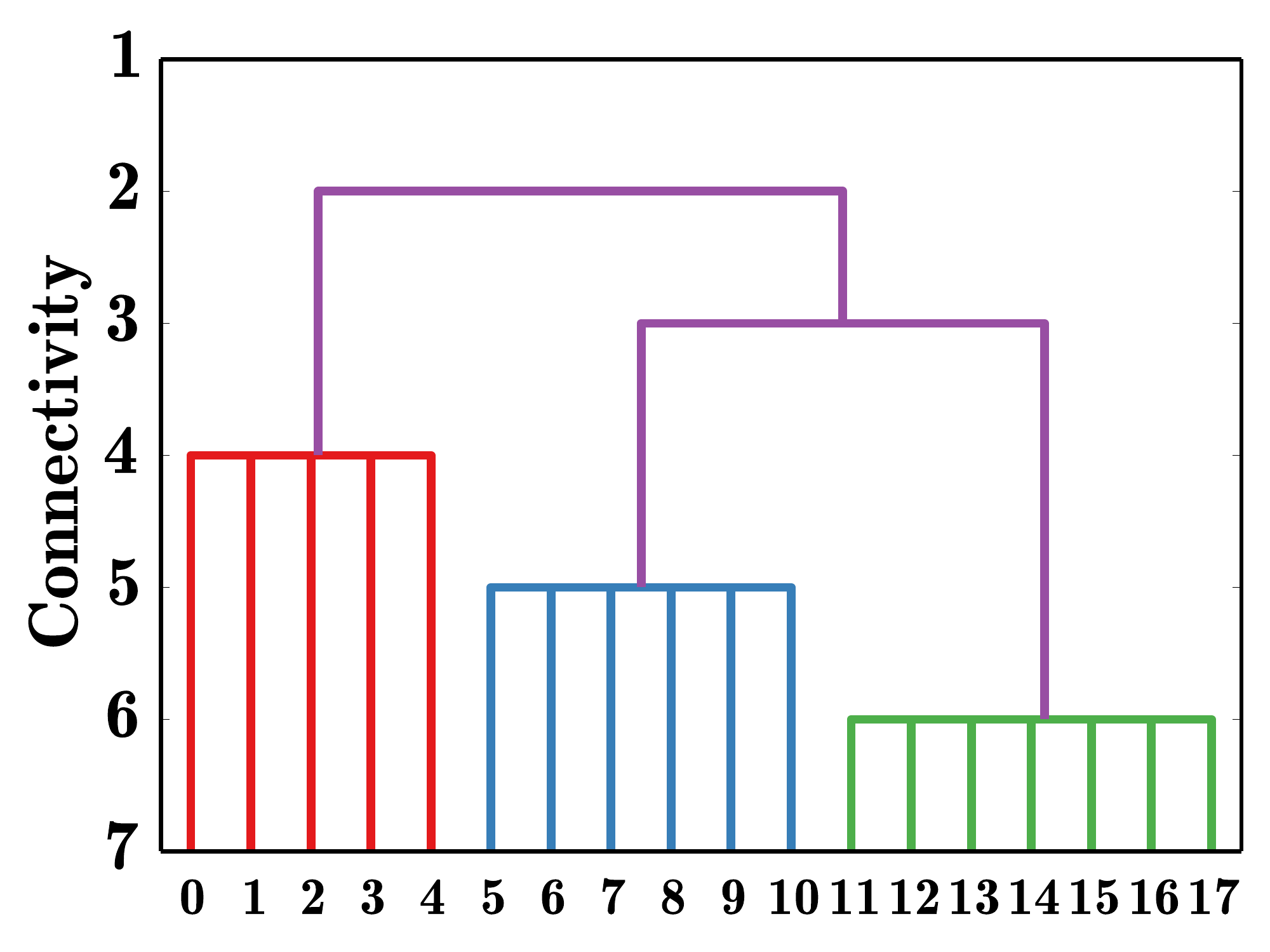}
}
\caption{An example of a graph and its connectivity dendrogram.}
\label{fig:dendrogram}
\end{figure}

As the connectivity can be considered to indicate the strength of a relationship,
we can define hierarchical clustering based on connectivity.
This can be visualized using a \emph{connectivity dendrogram}.
Figure~\ref{fig:dendrogram} shows a graph and its connectivity dendrogram.

The dendrogram of a cut tree can also be easily obtained.
Algorithm~\ref{alg:connectivity_dendrogram} explains the procedure,
which works in $O(n \log n)$ time.
Given a cut tree, it returns a tree $T' = (V'_T, E'_T)$ and a function $\ell$,
where each $X \in V'_T$ is a subset of the vertices of the original graph,
and $\ell(X)$ denotes the connectivity of vertex set $X$.
The underlying idea of the algorithm is
to look at edges in the cut tree in descending order of their weights
and merge the vertex subsets corresponding to both endpoints.
In experiments, this algorithm required 0.20 s and 0.30 s with the com-Orkut and indochina-2004 datasets, respectively
(excluding the time taken to construct the cut trees).

\begin{algorithm}[t!]
\small
\Procedure{\myfnsty{Connectivity-Dendrogram}$(T = (V_T, E_T, c_T))$}{
  \nl $\mathcal{V} \gets \set{\set{v} \mid v \in V_T}$;
  $V'_T \gets \set{\set{v} \mid v \in V_T}$; $E'_T \gets \emptyset$\;
  \nl $\ell(\set{v}) = \infty$ \textbf{for all} $v \in V_T$\;
  \nl \For{$e \in E_T$ in descending order of $c(e)$}{
    \nl $S, T \gets $ sets in $\mathcal{V}$ to which the ends of $e$ belong\;
    \nl $X \gets S \cup T$; $\mathcal{V} \gets (\mathcal{V} \setminus \set{S, T}) \cup \set{X}$\;
    \nl $V'_T \gets V'_T \cup \set{X}$;
    $E'_T \gets E'_T \cup \set{(S, X), (T, X)}$\;
    \nl $\ell(X) = c(e)$\;
  }
  \nl \Return{$T' = (V'_T, E'_T)$ and $\ell$}\;
}
\caption{Computing the connectivity dendrogram.}
\label{alg:connectivity_dendrogram}
\end{algorithm}

\section{Related Work}
\label{sec:related_work}

\myparagraphF{Graph Indexing Methods}
Because of their importance as back-ends for efficient network analysis,
graph indexing methods, i.e., methods that precompute and store some data structures to accelerate certain kinds of computation,
have been studied in the data mining community.
Examples include methods for
the point-to-point shortest-path distance~\cite{spq/exact/pll,spq/esa14,spq/approx/DasSarma2010}, %
single-source shortest-path distance~\cite{rel/sssp/kdd13,rel/sssp/sigmod12},
neighborhood function~\cite{ads/anf,ads/cohen15},
and personalized pagerank~\cite{Fujiwara2012}.
Cut trees can also be considered as an indexing method for graphs.

\myparagraph{Cut-Tree Construction Algorithms}
There has been little work on cut-tree construction algorithms,
other than the original algorithm of Gomory and Hu~\cite{cut-tree/gomory-hu}.
Gusfield's algorithm~\cite{cut-tree/gusfield} is very similar to the Gomory--Hu algorithm,
but does not include a contraction step.
Though slightly simpler than the Gomory--Hu algorithm,
preliminary experiments indicate that Gusfield’s algorithm is almost always slower for networks of interest.
Goldberg and Tsioutsiouliklis
proposed practical improvements to the Gomory--Hu algorithm
for instances arising from optimization problems~\cite{cut-tree/goldberg2001}.
Cohen \textit{et al.}\ studied thread-level parallelization
of the Gomory--Hu algorithm~\cite{cut-tree/parallel}.

\section{Conclusions}
\label{sec:conclusions}

In this paper, we have described a new algorithm for constructing cut trees from massive real-world graphs.
Our overall algorithm combines several new techniques covering
graph reduction, max-flow acceleration, and min-cut enumeration heuristics.
These techniques are tailored to real-world networks,
and, as confirmed by our experimental results, the resulting algorithm works surprisingly well.
Specifically, our algorithm constructed cut trees for web and social graphs with more than one billion edges,
some three orders of magnitude larger than can be handled by previous methods.
We also discussed some applications of cut trees to graph data mining.

\myparagraph{Repeatability}
As the implementations and datasets used in our experiments are available online,
our results are completely replicable.
The proposed method is available from \url{http://git.io/cut-tree}.
The previous methods are available from
\url{http://www.cs.princeton.edu/~kt/cut-tree/}
and \url{https://lemon.cs.elte.hu/trac/lemon}.
The datasets are available from
\url{http://snap.stanford.edu/data}
and \url{http://law.di.unimi.it/datasets.php}.

\myparagraph{Acknowledgement}
This work was supported by JSPS Grant-in-Aid for Research Activity Startup
(No. 15H06828) and JST, PRESTO.

{
\small
\bibliographystyle{abbrv}
\bibliography{main}

\begin{thebibliography}{10}

\bibitem{short}
T.~Akiba, Y.~Iwata, Y.~Sameshima, N.~Mizuno, and Y.~Yano.
\newblock Cut tree construction from massive graphs.
\newblock In {\em ICDM}, 2016.
\newblock to appear.

\bibitem{spq/exact/pll}
T.~Akiba, Y.~Iwata, and Y.~Yoshida.
\newblock Fast exact shortest-path distance queries on large networks by pruned
  landmark labeling.
\newblock In {\em SIGMOD}, pages 349--360, 2013.

\bibitem{mkecs/cikm13}
T.~Akiba, Y.~Iwata, and Y.~Yoshida.
\newblock Linear-time enumeration of maximal k-edge-connected subgraphs in
  large networks by random contraction.
\newblock In {\em CIKM}, pages 909--918, 2013.

\bibitem{asano2006mining}
Y.~Asano, T.~Nishizeki, M.~Toyoda, and M.~Kitsuregawa.
\newblock Mining communities on the web using a max-flow and a site-oriented
  framework.
\newblock {\em IEICE transactions on information and systems},
  89(10):2606--2615, 2006.

\bibitem{DBLP:journals/siamdm/Bang-JensenFJ95}
J.~Bang{-}Jensen, A.~Frank, and B.~Jackson.
\newblock Preserving and increasing local edge-connectivity in mixed graphs.
\newblock {\em {SIAM} J. Discrete Math.}, 8(2):155--178, 1995.

\bibitem{dataset/webgraph2}
P.~Boldi, M.~Rosa, M.~Santini, and S.~Vigna.
\newblock Layered label propagation: a multiresolution coordinate-free ordering
  for compressing social networks.
\newblock In {\em WWW}, pages 587--596, 2011.

\bibitem{dataset/webgraph1}
P.~Boldi and S.~Vigna.
\newblock The webgraph framework {I}: compression techniques.
\newblock In {\em WWW}, pages 595--602, 2004.

\bibitem{DBLP:journals/corr/BorassiN16}
M.~Borassi and E.~Natale.
\newblock {KADABRA} is an adaptive algorithm for betweenness via random
  approximation.
\newblock {\em CoRR}, abs/1604.08553, 2016.

\bibitem{mkecs/sigmod13}
L.~Chang, J.~X. Yu, L.~Qin, X.~Lin, C.~Liu, and W.~Liang.
\newblock Efficiently computing k-edge connected components via graph
  decomposition.
\newblock In {\em SIGMOD}, pages 205--216, 2013.

\bibitem{rel/sssp/sigmod12}
J.~Cheng, Y.~Ke, S.~Chu, and C.~Cheng.
\newblock Efficient processing of distance queries in large graphs: A vertex
  cover approach.
\newblock In {\em SIGMOD}, pages 457--468, 2012.

\bibitem{ads/cohen15}
E.~Cohen.
\newblock All-distances sketches, revisited: {HIP} estimators for massive
  graphs analysis.
\newblock {\em {IEEE} TKDE}, 27(9):2320--2334, 2015.

\bibitem{cut-tree/parallel}
J.~Cohen, L.~A. Rodrigues, and E.~P. Duarte~Jr.
\newblock A parallel implementation of gomory-hu's cut tree algorithm.
\newblock In {\em SBAC-PAD}, pages 124--131, 2012.

\bibitem{spq/approx/DasSarma2010}
A.~Das~Sarma, S.~Gollapudi, M.~Najork, and R.~Panigrahy.
\newblock A sketch-based distance oracle for web-scale graphs.
\newblock In {\em WSDM}, 2010.

\bibitem{spq/esa14}
D.~Delling, A.~V. Goldberg, T.~Pajor, and R.~F. Werneck.
\newblock Robust distance queries on massive networks.
\newblock In {\em ESA}, pages 321--333. 2014.

\bibitem{lemon}
B.~Dezs, A.~J\"{u}ttner, and P.~Kov\'{a}cs.
\newblock Lemon - an open source c++ graph template library.
\newblock {\em Electron. Notes Theor. Comput. Sci.}, 264(5):23--45, 2011.

\bibitem{1970:din}
E.~A. Dinic.
\newblock {A}lgorithm for {S}olution of a {P}roblem of {M}aximum {F}low in a
  {N}etwork with {P}ower {E}stimation.
\newblock {\em Soviet Math Doklady}, 11:1277--1280, 1970.

\bibitem{elias1956note}
P.~Elias, A.~Feinstein, and C.~E. Shannon.
\newblock A note on the maximum flow through a network.
\newblock {\em Information Theory, IRE Transactions on}, 2(4):117--119, 1956.

\bibitem{ford1956maximal}
L.~R. Ford and D.~R. Fulkerson.
\newblock Maximal flow through a network.
\newblock {\em Canadian journal of Mathematics}, 8(3):399--404, 1956.

\bibitem{Fujiwara2012}
Y.~Fujiwara, M.~Nakatsuji, T.~Yamamuro, H.~Shiokawa, and M.~Onizuka.
\newblock Efficient personalized pagerank with accuracy assurance.
\newblock In {\em KDD}, pages 15--23, 2012.

\bibitem{goldberg1984finding}
A.~V. Goldberg.
\newblock {\em Finding a maximum density subgraph}.
\newblock University of California Berkeley, CA, 1984.

\bibitem{DBLP:conf/esa/GoldbergHKTW11}
A.~V. Goldberg, S.~Hed, H.~Kaplan, R.~E. Tarjan, and R.~F.~F. Werneck.
\newblock Maximum flows by incremental breadth-first search.
\newblock In {\em Algorithms - {ESA} 2011 - 19th Annual European Symposium,
  Saarbr{\"{u}}cken, Germany, September 5-9, 2011. Proceedings}, pages
  457--468, 2011.

\bibitem{DBLP:journals/jacm/GoldbergT88}
A.~V. Goldberg and R.~E. Tarjan.
\newblock A new approach to the maximum-flow problem.
\newblock {\em J. {ACM}}, 35(4):921--940, 1988.

\bibitem{cut-tree/goldberg2001}
A.~V. Goldberg and K.~Tsioutsiouliklis.
\newblock Cut tree algorithms: an experimental study.
\newblock {\em Journal of Algorithms}, 38(1):51--83, 2001.

\bibitem{cut-tree/gomory-hu}
R.~E. Gomory and T.~C. Hu.
\newblock Multi-terminal network flows.
\newblock {\em Journal of the Society for Industrial and Applied Mathematics},
  9(4):551--570, 1961.

\bibitem{cut-tree/gusfield}
D.~Gusfield.
\newblock Very simple methods for all pairs network flow analysis.
\newblock {\em SIAM Journal on Computing}, 19(1):143--155, 1990.

\bibitem{hao1994faster}
J.~Hao and J.~B. Orlin.
\newblock A faster algorithm for finding the minimum cut in a directed graph.
\newblock {\em Journal of Algorithms}, 17(3):424--446, 1994.

\bibitem{dataset/snap}
J.~Leskovec and A.~Krevl.
\newblock {SNAP Datasets}: {Stanford} large network dataset collection.
\newblock \url{http://snap.stanford.edu/data}, 2014.

\bibitem{liben2007link}
D.~Liben-Nowell and J.~Kleinberg.
\newblock The link-prediction problem for social networks.
\newblock {\em Journal of the American society for information science and
  technology}, 58(7):1019--1031, 2007.

\bibitem{Orlin2013}
J.~B. Orlin.
\newblock Max flows in o(nm) time, or better.
\newblock In {\em STOC}, pages 765--774, 2013.

\bibitem{ads/anf}
C.~R. Palmer, P.~B. Gibbons, and C.~Faloutsos.
\newblock {ANF}: A fast and scalable tool for data mining in massive graphs.
\newblock In {\em KDD}, pages 81--90, 2002.

\bibitem{Qin2015}
L.~Qin, R.-H. Li, L.~Chang, and C.~Zhang.
\newblock Locally densest subgraph discovery.
\newblock In {\em KDD}, pages 965--974, 2015.

\bibitem{tarjan1974note}
R.~E. Tarjan.
\newblock A note on finding the bridges of a graph.
\newblock {\em Information Processing Letters}, 2(6):160--161, 1974.

\bibitem{Tsourakakis2015}
C.~Tsourakakis.
\newblock The k-clique densest subgraph problem.
\newblock In {\em WWW}, pages 1122--1132, 2015.

\bibitem{mkecs/edbt12}
R.~Zhou, C.~Liu, J.~X. Yu, W.~Liang, B.~Chen, and J.~Li.
\newblock Finding maximal $k$-edge-connected subgraphs from a large graph.
\newblock In {\em EDBT}, pages 480--491, 2012.

\bibitem{rel/sssp/kdd13}
A.~D. Zhu, X.~Xiao, S.~Wang, and W.~Lin.
\newblock Efficient single-source shortest path and distance queries on large
  graphs.
\newblock In {\em KDD}, pages 998--1006, 2013.

\end{thebibliography}
}

\end{document}